\documentstyle[12pt,aasms4]{article}
\lefthead{Gardini et al.}
\righthead{Cluster mass function in mixed models}

\def\be{\begin{equation}}
\def\ee{\end{equation}}
\def\bea{\begin{eqnarray}}
\def\eea{\end{eqnarray}}
\def\mincir{\raise -2.truept\hbox{\rlap{\hbox{$\sim$}}\raise5.truept
\hbox{$<$}\ }}
\def\magcir{\raise -4.truept\hbox{\rlap{\hbox{$\sim$}}\raise5.truept
\hbox{$>$}\ }}

\begin{document}
\title{CLUSTER CORRELATION IN MIXED MODELS}
\author{A. Gardini, S.A. Bonometto}
\affil{Dipartimento di Fisica G. Occhialini -- Universit\`a 
di Milano--Bicocca\\
INFN sezione di Milano -- Via Celoria 16, I20133 Milano, ITALY \\
e-mails: gardini@mi.infn.it -- bonometto@mi.infn.it
}
\authoremail{gardini@mi.infn.it}
\author{G. Murante}
\affil{Osservatorio Astronomico di Torino -- Pino Torinese\\
e-mail: giuseppe@to.astro.it
}
\authoremail{gardini@mi.infn.it}
\author{G. Yepes}
\affil{Departamento de F\'{\i}sica Te\'orica C--XI
-- Universidad Aut\'onoma de Madrid \\
Cantoblanco -- 28049 Madrid, SPAIN
 \\
e-mail: gustavo.yepes@uam.es
}
%\authoremail{gardini@mi.infn.it}

\begin{abstract}

We evaluate the dependence of the cluster correlation length $r_c$ on the
mean intercluster separation $D_c$, for three models with critical matter
density, vanishing vacuum energy ($\Lambda = 0$) and COBE normalized: 
a tilted CDM (tCDM) model ($n=0.8$) and two blue mixed models with
two light massive neutrinos yielding $\Omega_h = 0.26$ and 0.14
(MDM1 and MDM2, respectively). All models approach the observational
value of $\sigma_8$ (and, henceforth, the observed cluster abundance) and
are consistent with the observed abundance of Damped Lyman$\alpha$ systems.
Mixed models have a motivation in recent results of neutrino physics;
they also agree with the observed value of the ratio $\sigma_8/\sigma_{25}$,
yielding the spectral slope parameter $\Gamma$, and nicely fit LCRS
reconstructed spectra.

We use parallel AP3M simulations, performed in a wide box (side
360$\, h^{-1}$Mpc) and with high mass and distance resolution, enabling us
to build artificial samples of clusters, whose total number and mass range
allow to cover the same $D_c$ interval inspected through APM and Abell
cluster clustering data.

We find that the tCDM model performs substantially better than $n=1$ critical
density CDM models. Our main finding, however, is that mixed models
provide a surprisingly good fit of cluster clustering data.

\vskip 0.2 truecm


\end{abstract}


\section{Introduction}

The study of the clustering of galaxy clusters, in the early eighties, allowed
a basic advancement in our understanding of Large Scale Structure (LSS).
The discrepancy between the galaxy correlation length $r_g$ and the 
cluster correlation length $r_c$ (Bahcall \& Soneira 1983, Klypin
\& Kopylov 1983, but see also Hauser \& Peebles 1973) 
led to the introduction of the concept of bias (Kaiser 1984). 
Data on $r_c$ were then worked out, in further detail,
for Abell clusters by Peacock \& West (1992)
and Postman, Huchra \& Geller (1992), as well as for clusters in
APM and in Edinburgh--Durham Southern Galaxy catalogs, by
Dalton et al (1992), Nichol et al (1992) and Croft et al (1997). 

These analyses show that the value of $r_c$ depends on the 
mass threshold ($M_{th}$) of the cluster sample, through its mean 
intercluster separation $D_c =  n^{-1/3} (>M_{th})$, and that $r_c$
increases with $D_c$. However, $r_c$ values obtained from Abell and APM
data seem only partially consistent; this is to be partially
ascribed to different cluster definitions; Bahcall \& Burgett
(1986), Bahcall \& Cen (1992) and Bahcall \& West (1992)
suggested that observational ambiguities are wide enough to allow
to conjecture that the scaling relation $r_c \simeq 0.4\, D_c$ holds for 
$20 < D_c h/{\rm Mpc} < 100$. Herebelow, we shall refer to this relation
as BW conjecture. It ought to be born in mind that,
above $\sim 50\, h^{-1}$Mpc, such conjecture hinges on the estimates
of $r_c$ for 55 and 94$\, h^{-1}$Mpc mean separations, for richness $R
\geq 1$ and $R \geq 2$ Abell clusters, while APM data, for the
same $D_c$ range, give smaller $r_c$. Dekel et al (1989) and Sutherland
\& Efstathiou (1991) suggested that the projection effects and peculiar
inhomogeneities in the Abell sample might have biased upward $r_c$
at large $D_c$. Peacock \& West (1992), instead, confirmed such points
(see also Jing, Plionis \& Valdarnini, 1992).
Altogether, it may be fair to say that the controversy on the observational
behaviour of $r_c$ for high $D_c$ values has not been solved yet,
although, as we shall see, there may be good reasons to assess
that different cluster definitions play a key role.

This paper is devoted to a comparison of cluster clustering, as it emerges
from such data, with simulations of three cosmological models: a tilted CDM
(tCDM) model and two mixed models (MDM1 and MDM2) with cold+hot DM.
All models have critical matter density, vanishing vacuum energy,
and are COBE normalized. During the last few years, much attention
has been devoted to models with a positive cosmological constant $\Lambda$,
also because of the remarkable data sets concerning SN Ia (see,e.g.,
Riess et al 1998, Perlmutter et al 1998, and references therein).
In this work, we shall not debate whether mixed models can still offer
a fair fit to all cosmological data; they certainly do not fit SN Ia data,
unless their current interpretation was misled by some systematic bias.
In a number of cases, however, mixed models were just not tested and
the success of $\Lambda$--models to fit some data set was directly taken
as further evidence in their favour. In the case of cluster clustering,
we shall show that mixed models perform quite well and are surely
better than any other model with matter density parameter $\Omega_m = 1$
considered until now.

In order to fit cluster data with a model, a large simulation volume is
required; in fact, we need a fair sample of galaxy clusters for large $D_c$,
as well as adequate mass and force resolutions, to identify clusters
in a reliable way, for small $D_c$. Simulation parameters are therefore
set so to allow a sample of 90 clusters, at least, for large $D_c$ and
$\sim 60$ baryon--CDM particles per cluster, at least, for small $D_c$
(as we shall see, 60 particles correspond to $\sim 10^{14} h^{-1} M_\odot$).
Altogether, at redshift $z=0$, we shall therefore span a $D_c$ interval
ranging from $\sim 20$ to $80\, h^{-1}$Mpc.

Cluster clustering has been studied by various authors in simulations.
In particular, the behaviour of $r_c~vs.~D_c$, for standard CDM and
open CDM, was studied by Bahcall \& Cen (1992), Watanabe et al (1994),
Croft \& Efstathiou (1994), Eke et al (1996), Croft et al (1997),
Governato et al (1999).
Their results allow to conclude that CDM models with $n=1$ may
approach the observed behaviour of $r_c~vs.~D_c$, only for $\Omega_m < 1$.
The behaviour of $r_c~vs.~D_c$ in a mixed model was also
studied, using PM simulations, by Klypin \& Rhee (1994) and Walter
and Klypin (1996). Their work treated a different mix from those
considered here, using smaller box and resolution. Accordingly,
they could inspect only the $D_c$ interval running from
$\sim 20$ to 45$\, h^{-1}$Mpc. The behaviour they found
is only marginally consistent with a constant $r_c/D_c$ ratio, but
their model does not exhibit much improvement in respect to pure CDM.

The mixed models we consider here were selected on the basis of recent 
tests on $\nu$ flavour mixing, which seem to support a non--vanishing
$\nu$--mass. Mixing data come from the solar $\nu$ deficit
(see, e.g., Hampel et al 1996, for GALLEX, and Abdurashitov et al 1996,
for SAGE), the atmospheric $\nu$ anomaly (Fukuda et al 1994)
and the LSND experiment (Athanassopoulos et al 1995)
on $\nu$'s arising from $\mu^+$ and $\pi^+$ decay.
Barger, Weiler \& Whisnant (1998) and Sarkar (1999)
show that all above results can agree if a fourth sterile $\nu$ exists,
which can be however added without harming BBNS or LEP standard results.
Diagonalizing the mass matrix, they eventually obtain the four $\nu$--mass
eigenvalues, which split into two nearly degenerate pairs, corresponding to
$m_\nu \simeq 0$ and $m_\nu \sim 1.4$--1.5 eV.
It must be outlined that, within this picture, there remains no
contradiction among different experimental results, at variance with
earlier analyses which seemed to find contradictions between LSND and
other $\nu$--mixing results.

In a cosmological context, however, mixed models have been considered since
long. The transfer function for several mixed models was first computed
by Bonometto \& Valdarnini (1984). Results on mixed models were then found by
a number of authors (see, e.g., Achilli, Occhionero \& Scaramella 1985,
Valdarnini \& Bonometto 1985, Holtzmann 1989, for results obtainable
from the linear theory, and Davis et al 1992, Klypin et al 1993, 
Ghigna et al 1994, for early simulations). After the release
of LSND data, Primack et al (1995) performed simulations of models
with 2 massive $\nu$'s and yielding $\Omega_h = 0.20$ and found that
such mixture eased some problems met by greater $\Omega_h$ models.
The possibility of considering mixed models together with blue
spectra (primeval spectral index $n > 1$) was first considered by
Liddle et al (1996) and Lucchin et al (1996). In the former paper,
blue mixed models able to fit all linear and analytical constraints
were shown to exist. In the latter paper, inflationary models leading to
blue spectra were discussed and results of an N--body simulation of blue
mixed models were reported. Unfortunately, the model considered violated
some observational constraints. A systematic study of blue mixed 
models was recently performed by Bonometto \& Pierpaoli (1998) and
Pierpaoli \& Bonometto (1999), selecting those consistent
with CMB data and data predictable from the linear theory.

In the next section we show that the models considered here,
on the basis of $\nu$--physics motivation, are also suitable to
fulfill the main observational constraints. In {\S} ~3 we review the
technique used to simulate their non--linear evolution.
In {\S} ~4 we describe how clusters are selected in simulations.
Then, in {\S} ~5 we  describe how the 2--point correlation function 
and  its error estimates were worked out.
In {\S} 6, we will show the main results of the  $r_c ~vs.~D_c$ behaviour
derived from fits to the 2--point functions. {\S} 7 is devoted to
discussion of the results and the main conclusions we derived from this work.

\section{Model parameters}

The mixed models discussed in this paper were already considered
in a previous work by Gardini, Bonometto \& Murante (1999; hereafter Paper I)
They are models with two equal--mass massive $\nu$'s,
selected by requiring agreement with data which can be
fitted using the linear theory. More in detail, we required, first of all,
agreement with observations at top and bottom scales, i.e. with
COBE data and with the observed Damped Lyman$\alpha$ system (DLAS)
abundance (Storrie--Lombardi et al. 1995). Assuming 2 massive $\nu$'s
with $m_\nu \geq 1.5\, $eV, we adjusted the spectral index $n$ so to
agree with the above top and bottom data, choosing the minimum allowed value 
for the spectral amplitude ($A_\Psi$). Over intermediate scales,
the main constraints to be tested are at 8 and 25$\, h^{-1}$Mpc,
where we evaluated the mass variances $\sigma_{8,25}$ (see below).
From such values we can work out the expected spectral slope and cluster
abundance (again, see below); comparing their values with
observations we see that $m_\nu$ values up to $\sim 3\, $eV
can be considered, without violating such constraints.
Accordingly, we considered two values for the hot--dark--matter (HDM)
density parameter $\Omega_h$, yielding $m_\nu$ at the top and bottom
of the allowed interval.

A CDM model, selected so to fit the same data in a similar way,
was also studied, for the sake of comparison. While mixed models
require $n>1$, even with low $A_\Psi$, CDM may fit COBE data only if $n<1$.

Model parameters are shown in detail in Table I, while
fig.~1 shows the spectra obtained from the linear transfer
function $T(k)$ against APM reconstructed spectral points (\cite{Bau}).
The wavenumber $k$ is related to the comoving length scale
$L = 2\pi/k$ and to the mass scale $M = (4\pi/3)\rho_o L^3$,
where $\rho_o$ is the present density of the Universe. In Fig.~1 also
other data and results are shown, which will be discussed below.

\placetable{tab1}

The tCDM model approximates the APM galaxy spectrum slightly better than 
the standard CDM (sCDM), thanks to its increased slope. However,
it lays still quite below the spectral points around the peak at
$k \simeq 5 \times 10^{-2}h\, 
$Mpc$^{-1}$ (comoving length $\lambda \simeq 100$--120$\, h^{-1}$Mpc).
Blue mixed models, instead, show a stronger spectral
peak, occurring where $\nu$ free streaming bends a 
primeval spectrum steeper than Zel'dovich. This causes a steeper downward
spectrum for $5\, h^{-1}$Mpc$ \mincir \lambda \mincir 50\, h^{-1}$Mpc, which
might be related to the reasons why blue mixed models approach
the $r_c~vs.~D_c$ behaviour up to large $D_c$.

\begin{figure}
\begin{center}
\leavevmode
\plotone{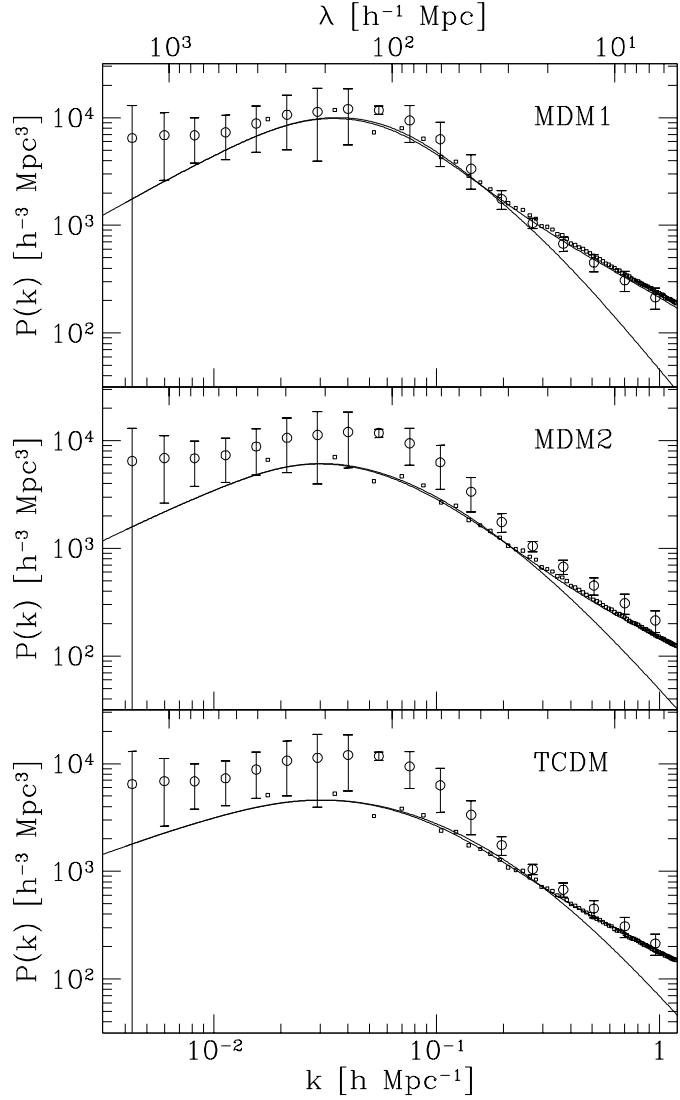}
\figcaption[Fig1.eps]{Spectra of the three models at $z=0$.
Solid curves give the linear power spectrum and the spectrum corrected 
for non--linearity, according to Peacock $\&$ Dodds (1996). Empty 
squares yield the
simulation spectra corrected for CIC (cloud--in--cell; see paper
I for more detail). Circles with
2$\, \sigma$ errorbars are the power spectrum measured from the APM survey.
\label{Fig1}}
\end{center}
\end{figure}

\placefigure{Fig1}

Mass variances are defined according to the relation
$$
\sigma^2 (L) = {\pi \over 9} \left(x_o \over L \right)^{n+3} A_\Psi
\int_0^\infty du u^{n+2} T^2 \left(u\over L\right) W^2 (u)~,
\eqno (1.1)
$$
with a top--hat window function $W(u) = 3(\sin u - u \cos u)/u^2$ and
$$
P(k) = {2\pi^3 \over 3} {A_\Psi \over x_o^3} (x_o k)^n~.
\eqno (1.2)
$$ 
Here $x_o$ is the comoving horizon distance. Using eq.~(1.1) we estimate the
parameter
$$
\Gamma = 7.13 \times 10^{-3} (\sigma_{ 8}/\sigma_{25})^{10/3}~,
\eqno (1.3)
$$
which, only for pure CDM models, is often approximated as $\Gamma 
\simeq \Omega_o h$ (see Efstathiou et al (1992). 
Peacock \& Dodds (1994), using APM data, and Borgani et al. (1994) 
using Abell/ACO samples, constrained $\Gamma$ within
the (2$\, \sigma$) intervals 0.19--0.27 and 0.18--0.25 respectively.
This parameter therefore tests the spectral slope above cluster mass scales.

Values of $\sigma_8$ are directly related to the expected
cluster number densities. A direct fit of data with simulations,
for the cluster mass function, is given in Paper I, where, however,
a different definition of cluster mass was used. This point will be
discussed again below and will be deepened in a forthcoming work.

\section{The simulations}

The three simulations considered here were already used in Paper I, 
to which we refer for details. They were performed using a parallel N--body 
code, based upon the serial public AP3M code of Couchman (1991), extended
in order to treat variable mass particle sets and used varying the time--steps,
when needed. We considered a box of side $L = 360\, h^{-1}$Mpc ($h$ is 
the Hubble parameters in units of 100 km${\rm \, s^{-1} Mpc^{-1}}$); 
here CDM+baryons were represented by $180^3$ particles, 
whose individual mass is $m_{180} = 2.22 \times 10^{12} h^{-1} M_\odot$ for 
tCDM. Mixed models also  involve 2 massive $\nu$'s with $m_\nu \simeq 
3.02\, $eV and  1.63$\, $eV, to yield $\Omega_h = 0.26$ and 0.14 (MDM1
and MDM2, respectively). Hence, {\it slow} particles, representing CDM+baryons,
have masses $\Omega_s\, m_{180}$ with $\Omega_s = \Omega_{CDM}+\Omega_{bar}
= 0.74$ and 0.86, respectively, while {\it fast} particles,
representing HDM, have masses ${1 \over 2}\Omega_h\, m_{180}$ (the ratio 2:1
between fast and slow particle numbers is required to set initial
conditions with locally vanishing linear momentum). Our force resolution
can be reported to a Plummer--equivalent smoothing parameter 
$\epsilon_{pl} \simeq 40.6\, h^{-1}$kpc. The comoving force and mass 
resolutions approach the limits of the computational resources of
the machine we used (an HP Exemplar SPP2000 X Class processor of the CILEA
consortium at Segrate--Milan).
The numerical resolution of our simulations were similar to other
simulations  of pure  CDM,  with different initial conditions performed
by  Colberg et al. (1997), Thomas et al. (1997), Cole et al. (1997),
and Governato et al (1999).
Mixed model simulations with a comparable dynamical range, instead,
have only been performed  by  Gross et al. (1998), but in a smaller volume.

\section{Cluster selection}

Different criteria can be used to select clusters in simulations.
In paper I, we identified clusters as virialized haloes.
Here we shall give results obtained with a cluster definition
aimed to approach more closely their observational definition. However,
we widely tested and compared results ensuing different definitions,
and a comprehensive discussion of the fit between particle sets
obtained with various criteria will be published elsewhere.
Let us however state that 2--point function outputs are fairly
robust; although specific values of the clustering length $r_c$,
at various $D_c$, have even significant variations, when the
cluster definition is changed, the general trend is always preserved:
the mixed models discussed here however fit observational data.
We shall illustrate this point with a few examples, without
giving outputs for whole sets of different cluster definitions.

In order to approach the observational pattern, here
clusters were found with a spherical overdensity (SO)
algorithm, based on a fixed sphere radius $R_a = 1.5\, h^{-1}$Mpc.
The details of the procedure are close to those suggested by
Croft \& Efstathiou (1994) and Klypin \& Rhee (1994). 
Hence, effects arising from the limiting magnitude of
(observational) samples, border effects and projection effects
are not included. (Also the sphere radius is fixed to mimic the Abell cluster
definition; clusters found in the APM survey were selected also with
smaller $R_a$. Our simulations were also used to test whether
systematic effects arise from a different choice of $R_a$;
the differences we found have no significant impact on the results that
will be shown here.)

More in detail, we start the procedure with a FoF algorithm,
which finds sets of $N$ CDM--baryon particles closer than $f$ times
the average inter--particle separation. Results reported here are obtained
using $f=0.2$ and $N=25$. Centers--of--mass (CM) of FoF groups are then
inspected as possible centers for SO. Starting from them, we follow an
iterative procedure: CDM--baryon particles within a distance $R_a$
from CM and their CM are found; this is repeated until we reach a stable
particle set and fix their CM. Only particle sets containing at least 25
particles are however kept. When, during the iterative procedure,
two spheres intersect, only the most massive particle set is kept.
Our procedure aims to find {\sl all} clusters above a
suitable mass scale. Loose requirements were therefore set on $f$, in
order to explore any possible matter condensation; the dependence of our
results on $N$ was also tested. Reducing $N$ obviously leads to more FoF
groups and a number of them survives the iteration procedure defined
hereabove. Most of such {\it extra--clusters}, however, do not contain
many particles. The result of such tests can be summarized by stating
that {\sl extra--clusters} of more than $\sim 60 $ particles, found lowering
$N$ down to 12, are less than $\sim 0.3\, \%$, in all cases; this percentage
has no further increase when still lower $N$ are taken. Henceforth, for
$N > 60$, i.e., for $M > 1.3 \Omega_s 10^{14} h^{-1} M_\odot \equiv M_{min}$,
our cluster samples can be considered complete. In fig.~2, we show the
relation between cluster masses and $D_c$ values, both at $z=0$ and at $z=0.8$.

\begin{figure}
\begin{center}
\leavevmode
\plotone{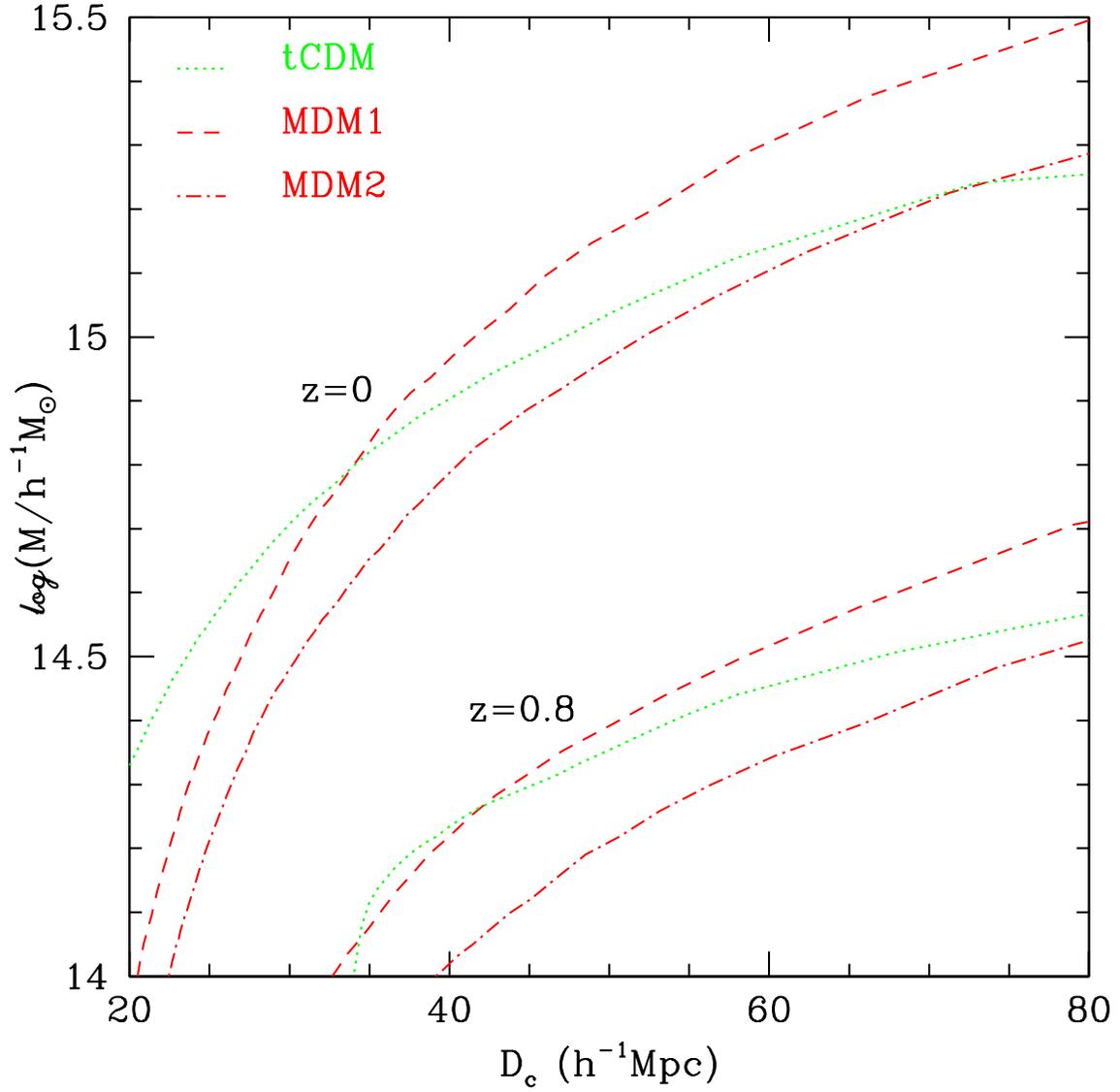}
\figcaption[Fig2.ps]{Mass limit ($M_{th}$) of cluster samples with mean
separation $D_c = n^{-1/3}(>M_{th})$.
\label{Fig2}}
\end{center}
\end{figure}

\placefigure{Fig2}

Among other tests, we also verified the size of virialized haloes
contained in clusters, as a function of their mass $M_c$.
Let $R_v$ be the radius encompassing a sphere, whereinside the
density contrast is 180, found starting from the CM of each cluster,
but whose actual center is attained through a suitable number of
iterations, so to be the CM of all particles within $R_v$ from it.
Let then $M_v$ be the mass of all CDM+baryon particles within $R_v$.
For large clusters, $R_v$ may exceed $R_a$; when $M_c$ is approximately
$< 7 \Omega_s \cdot 10^{14} h^{-1} M_\odot$, in general it is $M_v < M_c$.
In fig.~3 we show the values of $M_v$, as a function of $M_c$, for tCDM.
The trend is quite similar for mixed models. 

\begin{figure}
\begin{center}
\leavevmode
\plotone{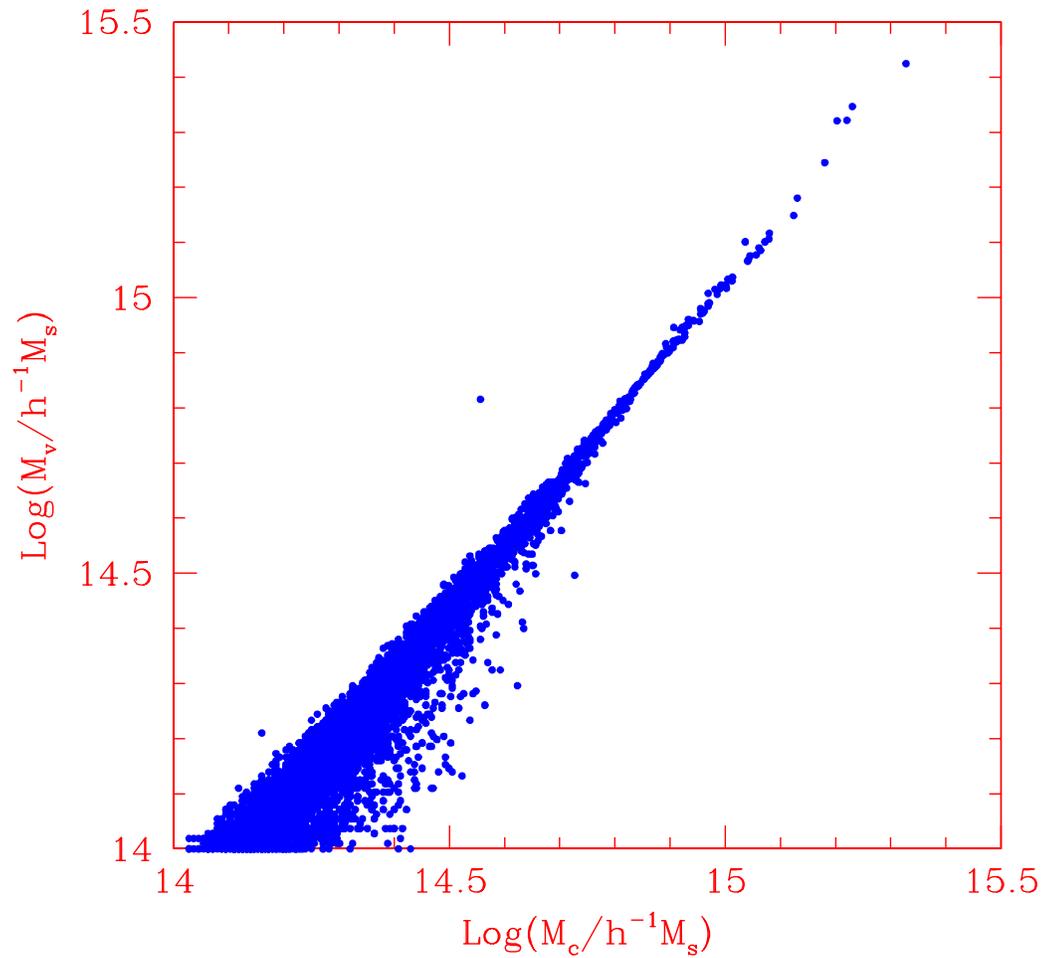}
\figcaption[Fig3.eps]{All clusters with $M_c > 10^{14} h^{-1} M_\odot$
are found to contain a virialized halo. Its mass $M_v$ is shown, as
a function of $M_c$, for the tCDM model. 
\label{Fig3}}
\end{center}
\end{figure}

\placefigure{Fig3}

Although $M_v$ tends to increase with $M_c$, the trend is clearly
not monotonic; this is due to the spread of the values we find for $R_v$
at any $M_c$. In spite of that, for $M > M_{min}$, all clusters contain
a virialized halo. However, if we order by mass the cluster set, using
$M_v$ instead of $M_c$, we find a different result. Hence, cluster sets,
whose mean distance is $D_c$, are different if we use $M_v$ instead of $M_c$.
It is then significant to compare the dependence of $r_c$ on $D_c$
for the two different orderings. In Paper I, clusters were given $M_v$
as mass; such definition is farther from observational criteria, but
is likely to be closer to physical requirements, e.g., if we aim
to compare simulation outputs with the expectations of a Press \& Schechter
approach. Herebelow, in a few cases, we shall test how the 
clustering length depends on $D_c$, when $M_v$ replaces $M_c$.
As we shall see, outputs depend significantly on the model, but
are substantially independent from the cluster definition.

\section{Cluster 2--point correlation function}

Using clusters in our simulation box, ordered according to their $M_c$,
we computed the 2--point correlation function $\xi(r)$, for a set of $D_c$
values, by applying the estimator
$$
\xi (r) = { D_c^6 N_{pairs} (r) \over L^3 \delta V(r)} - 1 ~.
\eqno (3.1)
$$
Here $ N_{pairs} (r)$ is the number of cluster pairs in the radial bin of
volume $\delta V(r)$, centered on $r$, and $L^3$ is the box volume.
Error bars for $\xi(r)$ were estimated
using the standard bootstrapping  procedure. (We checked the
convergence of the estimator of the standard deviation evaluated from
bootstrap realizations, by inspecting the third moment of the
bootstrap distribution. In all cases, convergence was attained when the
number of bootstrap realizations matched  the number of points in the
catalogues; see e.g. \cite{bradley})
We compared such errors with the usual Poisson errors, which were found
to be systematically smaller by a factor $\sim 2$.

We then performed two different fits to a power law 
$$
\xi(r)=(r/r_c)^{-\gamma} ~,
\eqno (3.2)
$$
over the distance range 5$\, h^{-1}$Mpc$< r < 25\, h^{-1}$Mpc: 
(i) A {\sl constrained} fit, assuming a constant
$\gamma = 1.8$. For the sake of example,
in fig.~4 we show such fit for $D_c = 30\, h^{-1}$Mpc. 

\begin{figure}
\begin{center}
\leavevmode
\plotone{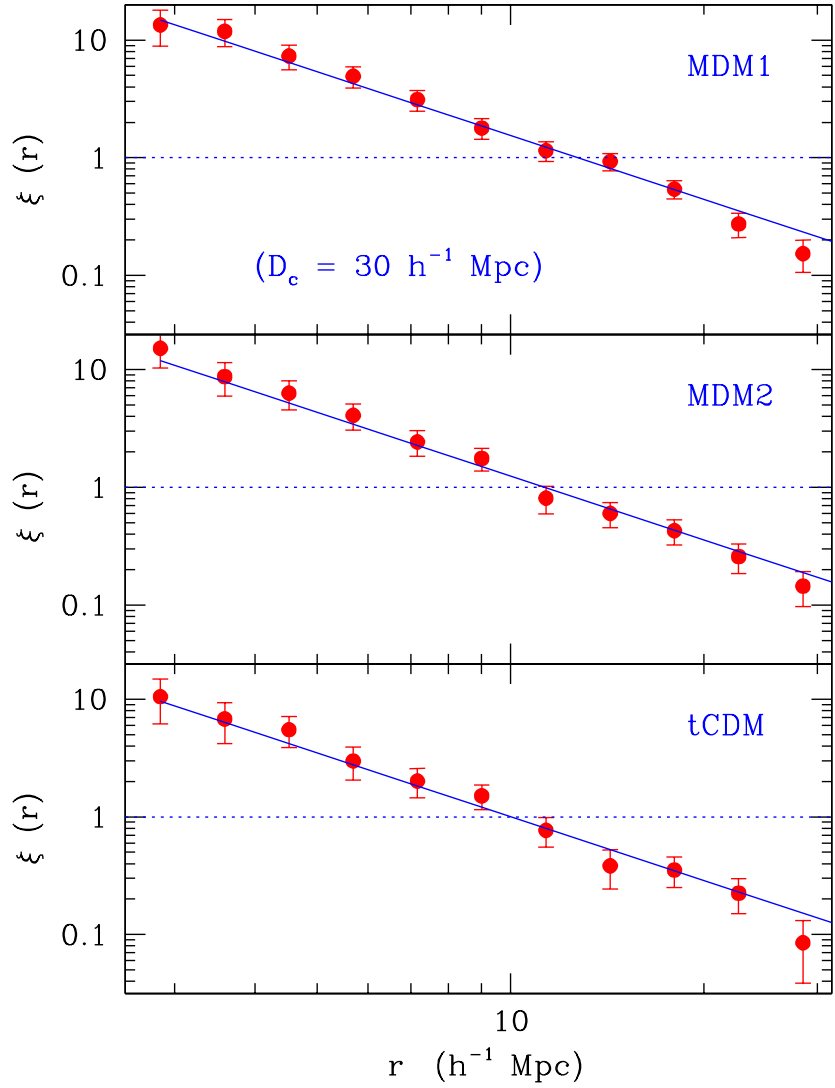}
\figcaption[Fig4.ps]{The 2--point correlation function estimate (solid points)
  and 1$\sigma$ bootstrap error bars for
$D_c = 30\, h^{-1}$ Mpc.
  A constrained fit with $\gamma = 1.8$ is also shown (solid line).
\label{Fig4}}
\end{center}
\end{figure}

\placefigure{Fig4}

(ii) An {\sl unconstrained} fit, 
allowing both $r_c$ and $\gamma$ to vary. Points were weighted by the
corresponding bootstrap errors and $r_c$ best--fit
values are also given with bootstrap errors. 
Such errors are obviously smaller for the constrained fit,
where our ignorance on $\gamma$ is hidden.
In general, large $D_c$ clusters yield best fit $\gamma$ values approaching
2, although 1.8 lays always within 1$\, \sigma$.
Our $r_c$ estimates are performed at
$z=0$ and $z=0.8$, to inspect cluster clustering evolution.

\section{Results}

In fig.~5 we report the $r_c ~vs.~D_c$ behaviour for tCDM, MDM1
and MDM2, for fixed $\gamma = 1.8$. In fig.~6 we give
results for the same cases, obtained with 2--parameters fits on $r_c$
and $\gamma$. Errors bars represent $1\, \sigma$ bootstrap errors
(see {\S} 3). Of course, error bars are smaller in the single parameter
fits, where our ignorance on $\gamma$ is hidden.

Together with the $r_c$ values obtained
from our simulations we also plot APM and Abell cluster data, the BW
conjecture, and the results from simulations performed by
Bahcall \& Cen (1992) and  Croft \& Efstathiou (1994).
Recent results obtained by Governato et al (1999), for a critical CDM model,
lay between the last two curves. Observational
points and error bars given in our figures were obtained from original work.
We draw the reader's attention on the fact that, in some recent work
studying cluster clustering in simulations, observational points and error
bars are not accurately reported.

\begin{figure}
\begin{center}
\leavevmode
\plotone{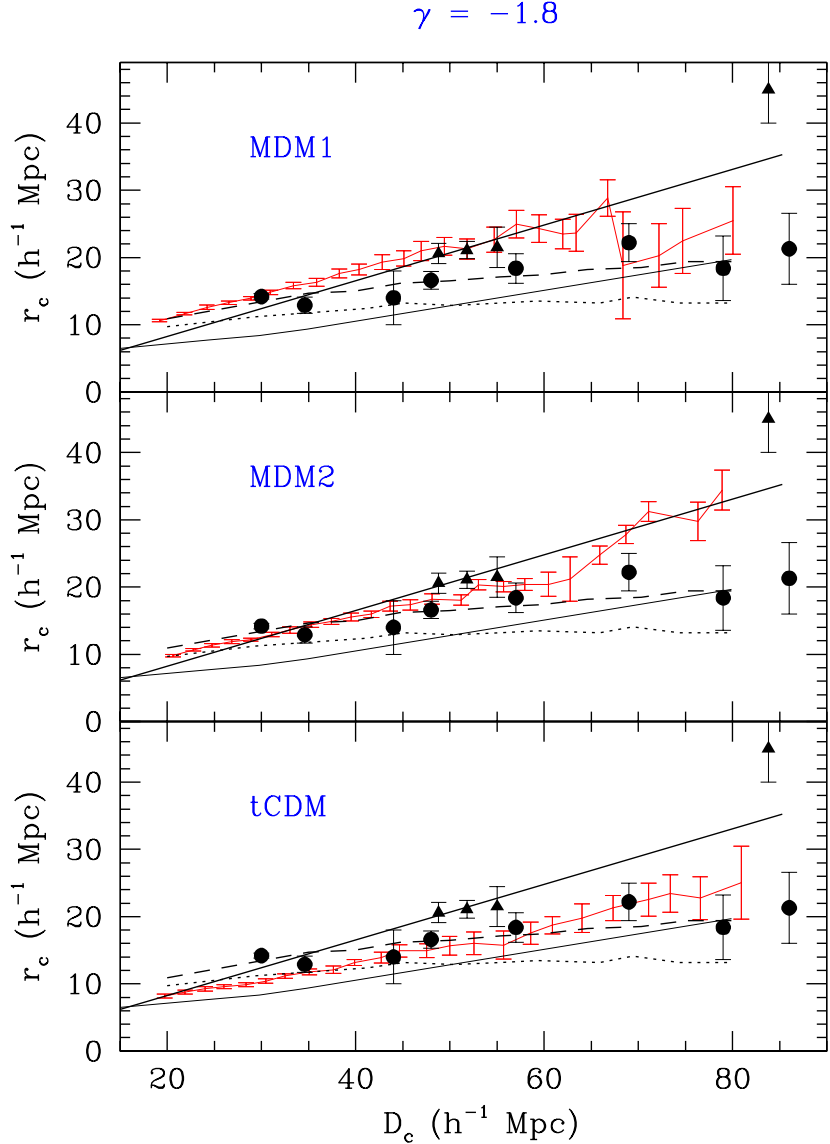}
\figcaption[Fig5.ps]{The cluster correlation length $r_c$ as a function of
~$D_c$, for the three models. Here $r_c$ is obtained by requiring 
$\gamma = 1.8$. 1$\sigma$ bootstrap error bars are also shown.  
$r_c$ values obtained from APM and Abell cluster data are also reported 
(filled circles and triangles, respectively). 
The thick solid line corresponds to  the BW conjecture. 
Results of SCDM simulation of Bahcall \& Cen (1992) are shown
by the thin solid line, while dashed and dotted lines refer to
Croft et al (1997) simulations of LCMD and SCDM models.
\label{Fig5}}
\end{center}
\end{figure}

\placefigure{Fig5}

\begin{figure}
\begin{center}
\leavevmode
\plotone{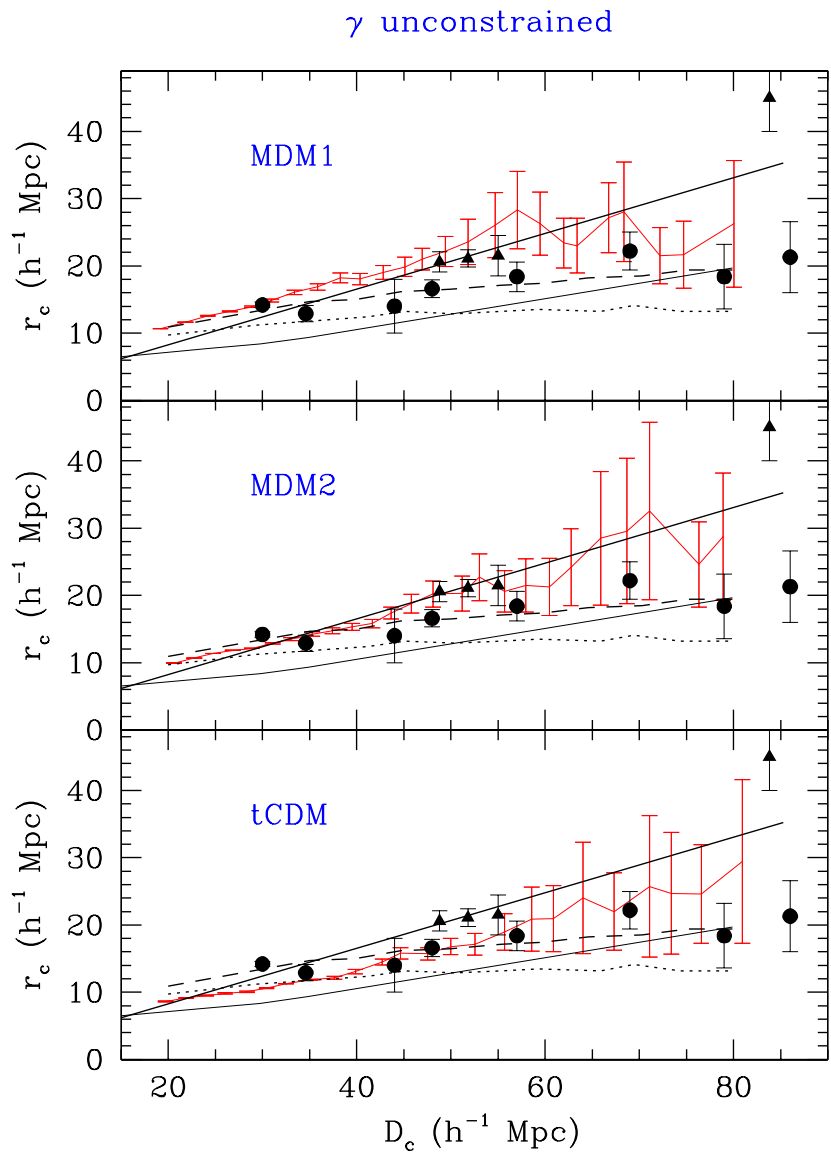}
\figcaption[Fig6.ps]{ Same as Fig \ref{Fig5} but with 
 $r_c$  obtained by simultaneously fitting it and $\gamma$ 
to the halo distribution. Error bars, representing  1$\sigma$ deviation derived
from the weighted least-square fits, are obviously larger  than  in 
the fixed--$\gamma$ case.
\label{Fig6}}
\end{center}
\end{figure}

\placefigure{Fig6}

A comparison of tCDM with data, shows that simulated and
APM data points are in fair agreement. In view of the better fits
obtainable with mixed models, shown in the same figures, one might
tend to overlook the improvement of tCDM in respect to CDM models with $n=1$,
which, instead, is significant. Our tCDM model, however, seems
to miss systematically the Abell catalog points and, thence, is far from
the BW conjecture, which tries to set a compromise between
Abell and APM results.

From this point of view, the performance of mixed models is better.
For low $D_c$, MDM1 tends to give $r_c$ values above the BW conjecture
(see, however, Lee \& Park 1999). A similar, but
less pronounced, effect exists also for MDM2. On intermediate scales
MDM1 sticks on the BW conjecture curve and meets two of the APM
points at the 2--$\sigma$ level only. MDM2, instead, seems to try
to compromise between APM and Abell points.
On top scales, the behaviours shown here by the two models are opposite.
The MDM1 behaviour, at such scales,
seems however somehow anomalous; such scales are those which
are most likely affected by cosmic variance and the MDM1
behaviour at $D_c > 65$--$70\, h^{-1}$Mpc should certainly be tested
with different model realizations. Furthermore,
unconstrained fits tend to indicate that such
discrepancy arises from different correlation function slopes.

It may also be significant to consider the unconstrained fit obtained
ordering clusters according to $M_v$ masses, which is shown in fig.~7.
Let us notice that: (i) in most cases, error bars are smaller;
(ii) the peculiar feature for MDM1 at large $D_c$ has disappeared.
It is likely that such improvement is related to a more direct
physical significance of the mass $M_v$ and the (variable) radius $R_v$.
Taking a fixed radius $R_a$, instead, risks to accentuate a dependence
on local peculiar features. In principle, this is more likely to
occur for small mass clusters, for which significant volumes, still
unaffected by virialization processes, lay within $R_a$. In our
simulation volume, however, we have a large deal of low--mass clusters
and this allows an efficient averaging over local realizations.
At the top mass end, instead, the sample is more restricted and 
we must mostly rely on the virialization process, rather than on
sample averaging, to smear off local peculiarities. Our results
seem to indicate that significant memory of initial conditions
is kept also below $R_v$.

\begin{figure}
\begin{center}
\leavevmode
\plotone{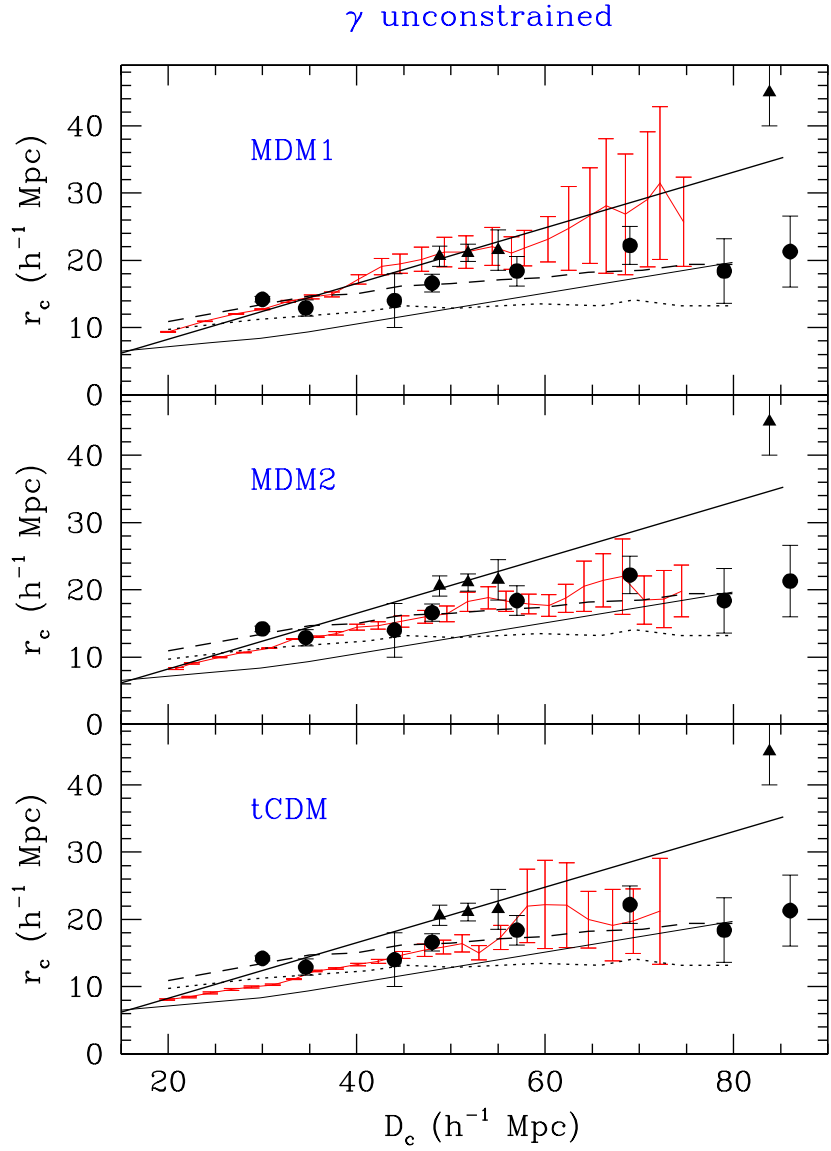}
\figcaption[Fig7.eps]{ Same as Fig \ref{Fig6} but  with 
 $D_c$  obtained ordering clusters according to $M_v$.
\label{Fig7}}
\end{center}
\end{figure}

\placefigure{Fig7}

\begin{figure}
\begin{center}
\leavevmode
\plotone{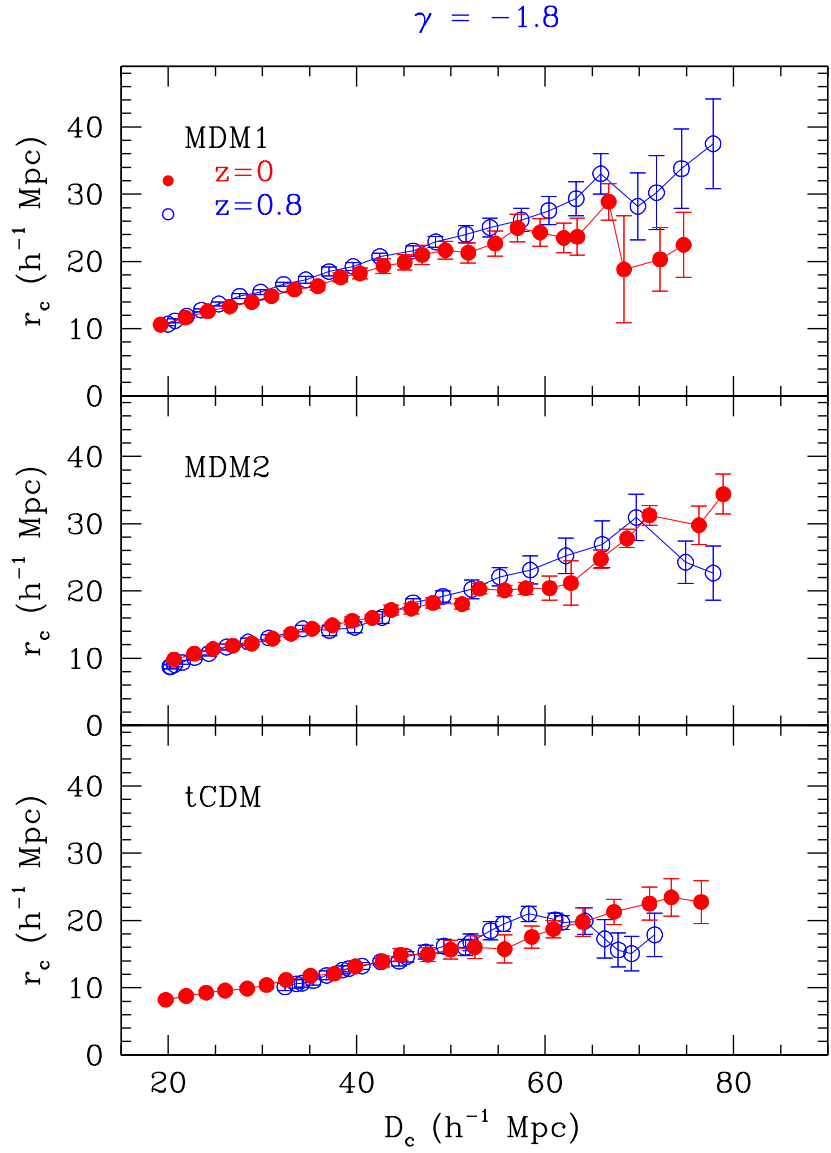}
\figcaption[Fig8.ps]{Comoving clustering evolution from $z=0.8$ to $z=0$
  obtained by fitting   $r_c$ with $\gamma=1.8$.
\label{Fig8}}
\end{center}
\end{figure}

\begin{figure}
\begin{center}
\leavevmode
\plotone{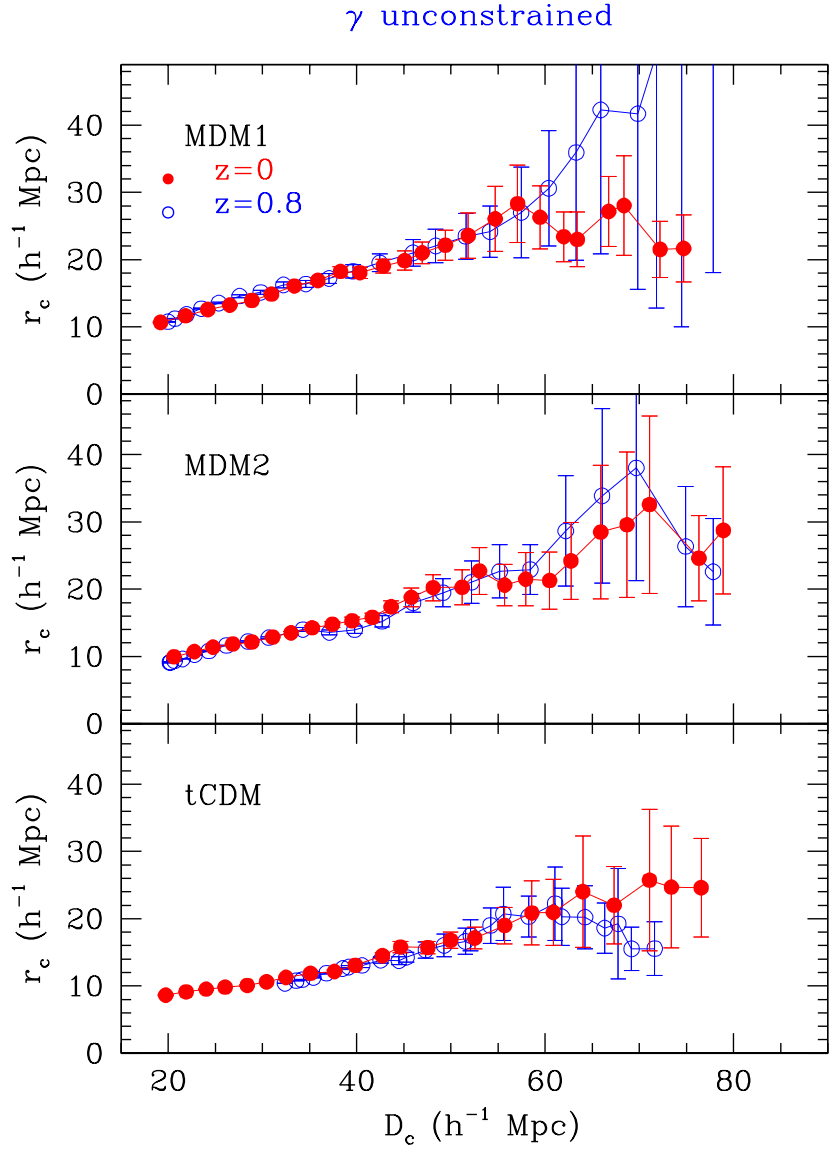}
\figcaption[Fig9.ps]{Same as Fig \ref{Fig8}, but from simultaneous fits 
 of $r_c$ and $\gamma $. MDM1 shows some anomaly.
\label{Fig9}}
\end{center}
\end{figure}

In Fig.s \ref{Fig8} and \ref{Fig9} we report a comparison between the 2--point
function results at $z=0$ and $z=0.8$, obtained using $M_c$.

\placefigure{Fig8}

\placefigure{Fig9}

Unconstrained fits at $z=0.8$
are rather noisy at large $D_c$, in particular for top scales.
Constrained fits, instead, might be taken as an indication of
clustering evolution on top scales. Here, perhaps, there is a
further evidence of anomaly in MDM1, which is the only case when
clustering seems however weeker at $z=0$ than at $z=0.8$.
Apparently, all models seem to indicate a greater clustering
length at scales between 50 and 65--70$\, h^{-1}$Mpc for $z=0.8$.
Apart of MDM1, instead, this is inverted above 70$\, h^{-1}$Mpc.

\begin{figure}
\begin{center}
\leavevmode
\plotone{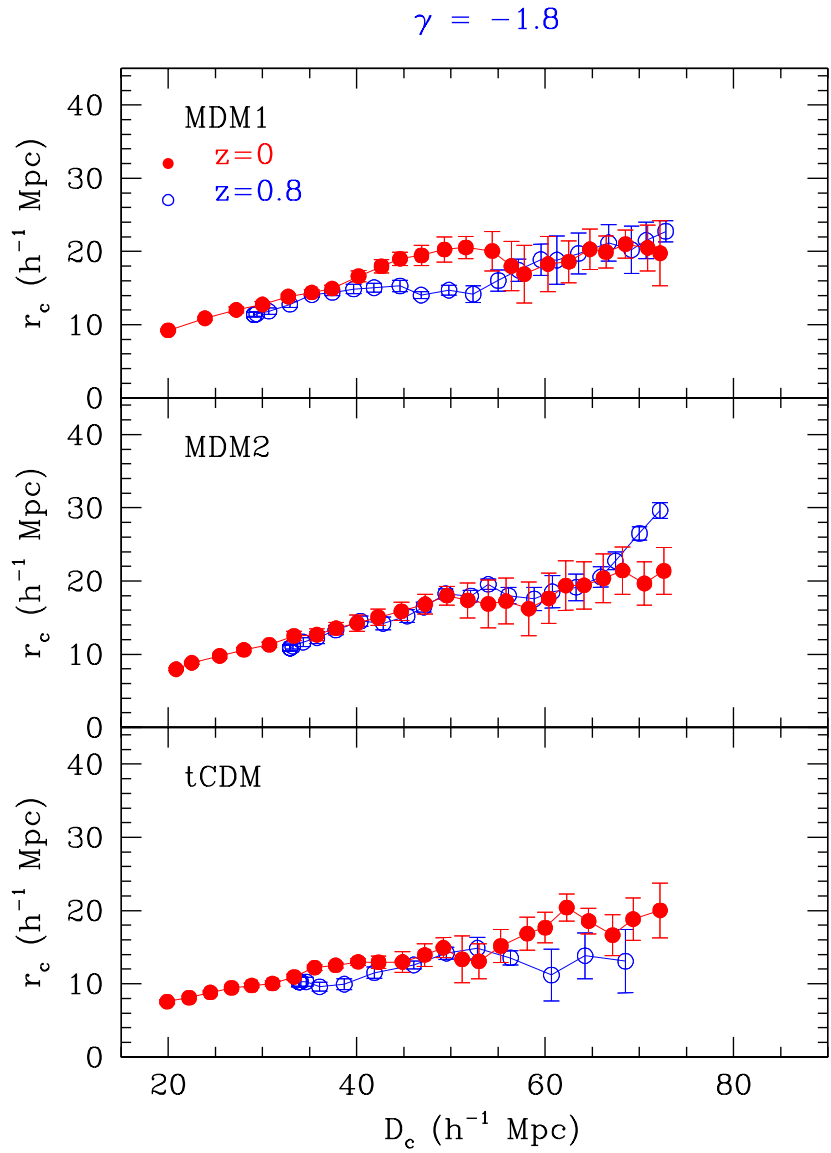}
\figcaption[Fig10.eps]{The same as Fig \ref{Fig8}, but $D_c$ are obtained 
ordering clusters according to $M_v$. Plots are less noisy,
when such mass determination is used.
\label{Fig10}}
\end{center}
\end{figure}

\placefigure{Fig10}

Fig \ref{Fig10}, instead, shows results on cluster evolution based on
$M_v$ ordering and using constrained fits. The kind of evolution
found hereabove seems confirmed, while MDM1 anomaly is reduced.

\section{Discussion}

Previous numerical results on cluster clustering, based on models
with $\Omega_m = 1$, gave a behaviour of  $r_c~vs.~D_c$ a few $\sigma$'s
below observational results. The only exception are Bahcall \&
Cen (1992), whose numerical study involves peculiar extrapolations and
however succeeds to meet two APM points at large $D_c$ only.
When considering subcritical CDM models (OCDM), the same authors
obtain a behaviour close to the BW conjecture. However, this
is not fully confirmed by later numerical studies; although they clearly
indicate that, in OCDM models, the $D_c$ dependence on $r_c$
is consistent with APM points, Abell cluster points
seem to require a still steeper dependence than in OCDM.
Such findings, however, were currently interpreted
as an indication that observational data on cluster clustering may be
approached only by models with $\Omega_m < 1$ and led to arguing that
the observed dependence of $r_c$ on $M_{th}$ is somehow related to an
early cluster formation.

The behaviour we find for tCDM does not support such kind of inference.
Taking a spectral index $n \neq 1$ has no substantial effect on
the time of cluster formation, which is quite similar to standard CDM.
In view of the more stricking outputs for mixed models, we must
not disregard the result we find for tCDM. The
$r_c~vs.~D_c$ behaviour of clusters in such model is analogous
to previous outputs for OCDM and lays well above the very
behaviour obtained by Bahcall \& Cen (1992) for standard CDM.
Our findings are that, taking $n < 1$, the clustering of clusters in
an $\Omega_m = 1$ model approaches the behaviour obtained from APM cluster
data.

Even more striking is the cluster clustering behaviour for the mixed models
considered in this work, which are based on low mass $\nu$'s.
In this case, the slope of the $r_c~vs.~D_c$ behaviour approaches
the BW conjecture and, more significantly, within
1 or 2--$\sigma$ bootstrap error bars, we mostly find
consistency both with APM and Abell results. The only exception
could be the point of Abell clusters of richness $R > 2$,
which is approached only by MDM2. Our results, however, support
previous claims that, at the top mass end, wider samples
may be required to suppress the cosmic variance. 

Quite in general, it can be reasonable to consider cluster clustering
as a measure of the spectral power on scales exceeding $\sim 25\, h^{-1} $Mpc.
All models with $\Omega_m = 1$ ought to have similar values of $\sigma_8$,
in order to be consistent with the observed cluster abundance.
Accordingly, the power at scales $\sim 25\, h^{-1} $Mpc is basically gauged
by the value of the $\Gamma$ parameter. It may not be a case that
models perform in a better way as their $\Gamma$'s approach the
observational interval. 

Surely, in the case of mixed models, the
situation is complicated by the presence of a hot component, which
may slow down the gravitational growth in the non--linear regime,
in a scale--dependent fashion.
Previous results for mixed models concerned a mix including 30$\ \%$
of HDM, due to a single $\nu$ with mass $m_\nu \simeq 7\, $eV. Here we
deal with $\nu$'s 3--4 times lighter; accordingly, when cluster formation
begins, their speeds are $\sim 3$--4 times greater. Peculiar effects
of MDM are therefore significantly reinforced. Hence,
besides of having a suitable spectral slope, the mixed models treated
here are still more different from standard CDM, because of the
late $\nu$ derelativisation.

In this paper we showed that critical CDM models, with a blue
spectrum suitably ``compensated'' by a light--$\nu$ component,
besides of fitting most LSS and CMB data, are able to follow
APM and Abell cluster clustering data. This result  adds 
to those discussed in Paper I, where critical blue mixed models were shown
to provide a good fit to  the cluster mass
function and to be in agreement with Donahue at al. (1998) findings
concerning high--$z$ cluster abundance.

\vfill\eject

\newpage

%%%%%% TABLES 

\begin{table}
\caption{
Parameters of the models; besides of
input parameters or quantities derived  from the
linear theory we report the value of $N_{cl}$ (sim) (number of
clusters of mass $> 4.2 \times 10^{14} h^{-1} M_\odot$ in a box
side of $100 h^{-1}$Mpc) as obtained in Paper I.
The normalization to COBE quadrupole was deliberately kept at the
$\sim 3 \, \sigma$ lower limit, in order leave some room to the
contribution of tensor modes, but keeping however consistent with data.
The expected interval for $N_{cl} $ is 4--6, but models with
$N_{cl}$ up to 8--10 cannot be safely rejected. 
$L_\alpha \equiv \Omega_b \Omega_{\rm coll}(z=4.25,M=5\times 10^{9} h^{-1}
M_\odot) \times 10^{3} $  accounts for the amount 
of gas expected in Damped Lyman--$\alpha$ systems. More details can be found
in Paper I. Data provided by Storrie--Lombardi et al. (1995)
give $L_\alpha > 2.2 \pm 0.6$. This is one of the most stringent
indicators of the model capacity to produce high--z objects.
\label{tab1}
}
\begin{center}
\begin{tabular}{l c c c}
\tableline \tableline
\multicolumn{4}{c}{} \\
%\multicolumn{4}{c}{ Model data }  \\
&\multicolumn{1}{c}{MDM1}&\multicolumn{1}{c}{MDM2}&\multicolumn{1}{c}{TCDM}
\\
\tableline
 $\Omega_h$           &          0.26     &      0.14    &     ---- \\
 $m_\nu/$eV           &         3.022     &     1.627    &     ---- \\
 $\Omega_b \cdot 10^2$ &         6.8      &       9      &      6   \\
 $n$                  &          1.2      &     1.05     &     0.8  \\
 $Q_{PS,rms}/\mu$K    &         12.1      &      13      &    17.4  \\
 $\sigma_8$           &         0.75      &    0.62      &    0.61  \\
 $\Gamma$             &         0.18      &    0.23      &   0.32   \\
 $N_{cl}$ (PS; $\delta_c = 1.69$)  &        14.      &      5.2      &     5.7\\
 $N_{cl}$ (sim)       &          10.      &     4.7      &    6.0   \\
 $L_\alpha $      &          1.3      &     1.2      &    1.3   \\
%\multicolumn{4}{c}{} \\
\multicolumn{4}{c}{} \\
\tableline \tableline
\end{tabular}
\end{center}
\end{table}

\newpage

%%% Figure Captions

\end{document}